\documentclass[12pt]{article}
\usepackage{fullpage}
\usepackage{ulem}
\usepackage{graphicx}
\usepackage{amssymb}
\usepackage{amsmath}
\usepackage{multicol}
\usepackage{ulem}
\newcommand{\fg}{Figure }
\newcommand{\Fig}{Figure }
\newcommand{\pr}{\partial{}}

\numberwithin{equation}{section}
\newcommand{\bphi}{\mbox{\boldmath $\phi$}}
\newcommand{\bx}{\mbox{\boldmath $x$}}

\newcommand{\skf}{Skyrme-Faddeev }
\newcommand{\news}{\setcounter{equation}{0}\quad}

\title{
\vskip 2cm \Huge Massive Hopfions.}
\author{David Foster\\[10pt]
 School of Theoretical Physics, \\
Dublin Institute for Advanced Studies, \\
10 Burlington Road, Dublin 4, Ireland.\\
{\normalsize email:dfoster@stp.dias.ie}}
\date{    }

\begin{document}
\maketitle
\begin{abstract}
 The \skf model is a $(3+1)$-dimensional model which has knotted, string-like, soliton solutions. In this paper we investigate a \skf model with an $SO(3)$ symmetry breaking potential. We then rescale this model and take the mass to infinity. This infinite mass model is found to have compact knotted solutions. In all of the investigated massive models we find similar charged solutions as in the usual, $m=0$, model. We also find that their energies follow a similar $E \sim Q^{3/4}$ power growth as the $m=0$ model.  
\end{abstract}
\newpage
\section{Introduction.}\news
The \skf model is a $(3+1)$-dimension modified $O(3)$-sigma model \cite{fadd}. This theory has finite-energy string-like solutions and has a potential application as a low energy effective theory of QCD \cite{hopf-QCD}. The theory usually has two components, a sigma term which is a quadratic of derivatives and a Skyrme term which is quartic in derivatives. These are the minimum terms required to provide a scale for Derricks theorem, \cite{Derrick}, in $(3+1)$-dimensional flat space. The theory is described by the Lagrangian
\begin{equation}
 L = \frac{1}{32 \pi^2 \sqrt{2}} \int \left(\pr_\mu \bphi \cdot \pr^\mu \bphi -\frac{1}{2} (\pr_\mu \bphi \times \pr_\nu \bphi)\cdot(\pr^\mu \bphi \times \pr^\nu \bphi) - 2m^2 V(|\bphi|) \right)d^3x, \label{L}
\end{equation}
where $\bphi$ is a three component unit vector, $\bphi = (\phi_1,\phi_2,\phi_3)$. At fixed time $\bphi$ is the map $\bphi:\mathbb{R}^3 \to S^2$. A condition for finite energy solutions  of \eqref{L} is that the field must tend to a constant value at spatial infinity, which we select to be $\bphi(t,r=\infty) = (0,0,1)=\mbox{\textbf{e}}_3$ for all time, $t$. This allows a one-point compactification of the domain, $\mathbb{R}^3 \cup \{0\} \sim S^3$. So static, finite-energy, solutions are a map 
\begin{equation}
\bphi:S^3 \to S^2. \label{finite-map}
\end{equation}
This map,\eqref{finite-map}, belongs to an equivalence class characterised by the homotopy group $\pi_3 (S^2)=\mathbb{Z}$. This shows that there is an integer topological invariant associated with $\bphi$, known as the Hopf invariant. In this case we will refer to the Hopf invariant as the topological charge $Q$. \\
The topological charge can be found by firstly defining an area form $\omega$ on $S^2$. Then we can define $g=\bphi^\ast \omega$, which is the pull-back of $\omega$ by $\bphi$. 
Due to the second cohomology group of the three sphere being trivial, $H^2(S^3) =0$, all closed forms on $S^3$ are exact forms. Therefore, we can now re-define the exact form $g$ by the $1$-form $a$ as $g=da$. Therefore we can express the topological charge, $Q$, as
\begin{equation}
 Q = \frac{1}{4 \pi} \int_{S^3} g\wedge a. \label{non-local-charge}
\end{equation}
This non-local definition of the topological charge, \eqref{non-local-charge}, is not particularly useful in this context. Instead the topological charge of \eqref{finite-map} can be found as the linking number of loops in the domain. These loops are formed as preimages of two distinct points on the target space. For example if we define the two curves $C_p$ and $C_q$ as the preimages of the points $\mbox{\textbf{p}}$ and $\mbox{\textbf{q}}$. If we then choose a smooth surface $D$, with a boundary $C_p$, the linking is defined as
\begin{equation}
 \mbox{link}(C_p,C_q) = \sum_{D \cap C_q} \pm 1, \nonumber
\end{equation}
where the $\pm$ refers to the relative orientations of $D$ and $C_q$. This definition of the topological invariant can be shown to be the same as \eqref{non-local-charge}, \cite{Bott}.\\
There is a well known lower bound on the energy, \cite{Vak,ward-sk,lin-yang,kung-yu}, which is based on an involved argument using Sobolev-type inequalities,
\begin{equation}
 E \geq c~Q^{3/4} ~ \mbox{where} ~ c=\left(\frac{3}{16} \right)^{3/8}. \label{Hopf-bound}
\end{equation}
The fractional power of \eqref{Hopf-bound} is proven to be optimal, but the value of $c$ might not be. Ward, \cite{ward-sk}, was motivated by his study of the \skf theory on a unit three-sphere to propose that $c=1$ might be a more optimal value. \\
The Hopfion location is commonly defined as the curve $C=\bphi^{-1}(0,0,-1)\equiv \bphi^{-1}\mbox{\textbf{e}}_{-3}$, which is the antipodal value to the boundary vacuum value. \\
For $m=0$ there have been many extensive and detailed investigations into the static minimum energy solutions of \eqref{L}, \cite{fadd,battye-sutcliffe,ward,glad,ward-latt,heit}. For charges one to seven it is believed their respective global minimum energy solutions have all been identified. 
It is known \cite{sutcliffe-good} for topological charges one to three that the minimum energy solutions have a planar loop location curve. Topological charges five and six have the minimum energy solutions of two linked Hopfions \cite{sutcliffe-good}. For topological charge seven the minimum energy location curve is a trefoil knot \cite{sutcliffe-good}. In \cite{sutcliffe-good} Sutcliffe devised a new knotted rational map ansatz as initial conditions. These initial conditions were then energetically minimised to give new minimum energy solution candidates for a large class of topological charges. We shall describe and make use of this technique later.

\section{$m>0$.}
The actual model of interest here is a modification of the usual \skf model. It is modified by an additional potential term, so $m>0$ in \eqref{L} and
\begin{equation}
 V(\bphi)=(1-\phi_3). \label{pot}
\end{equation}
This choice of potential, \eqref{pot}, is not general but is one of the simplest choices. A potential term, which breaks global $SO(3)$ symmetry, is also generated on derivation of the theory \eqref{L} as an infrared limit of $(3+1)$-dimensional $SU(2)$ Yang-Mills theory \cite{L-F-A-N}. If we restrict ourself to the plane this model reduces to the old Baby Skyrmion model. Also this potential, \eqref{pot}, meets the finite energy criteria for the chosen boundary condition; where the single vacuum of $V(\bphi)$ is also the chosen boundary value. The potential, \eqref{pot}, increases the energy density along the location of the Hopfion. This is because the location curve, $C=\bphi^{-1}(0,0,-1)$, corresponds to the maximum of the potential \eqref{pot}. Therefore, we expect the Hopfion location curve to become smaller in length for increasing $m$. We also expect the Hopfion string to become finer with increasing $m$; as it is analogous with the Baby Skyrmion model. This is best understood by an asymptotic analysis where we approximate the field for large $r$ as
\begin{equation}
 \bphi = (\epsilon_1,\epsilon_2,1-\epsilon_1^2-\epsilon_2^2) +\mathcal{O}(\epsilon_a^3), \nonumber
\end{equation}
where $a\in(0,1)$. For large $r$ we know that, due to finite energy criteria, $|\pr_i \epsilon_a| <1$ therefore the energy density associated with \eqref{L} becomes
\begin{equation}
 \mathcal{E} = (\pr_i \epsilon_a)^2 +2 m^2 \epsilon_a^2 + \mathcal{O}\left( (\pr_i \epsilon_a)^4\right). \nonumber
\end{equation}
 Where $\epsilon_a$ is a solution of the partial differential equation
\begin{equation}
 (\Delta - 2 m^2) \epsilon_a =0. \label{asym-e}
\end{equation}
Separating $\epsilon_a$ into radial and angular components as $\epsilon_a =r^{-\frac{1}{2}}R_a(r)\Theta_a(\theta,\psi)$. Where $R_a(r)$ is a purely radial function, $\Theta_a$ is a spherical harmonic $(\nabla_{s^2} \Theta_a = -\lambda(\lambda+1) \Theta_a)$ and $(r,\theta,\psi)$ are the usual spherical polar coordinates. Then solving \eqref{asym-e} we gain an asymptotic approximation for $R_a(r>>1)$,
\begin{equation}
 R_a(r>>1) \sim C_a \sqrt{\frac{\pi}{2 \sqrt{2}m r}} e^{-\sqrt{2}m r}\left(1 + \mathcal{O} \left(\frac{1}{\sqrt{2}mr}\right) \right). \nonumber
\end{equation}
%full working out is in the bottom draw of university desk
Where $C_a$ is a constant of integration. This shows that Hopfions become increasingly exponentially located as $m$ increases. Hence the Hopfions in this massive theory will have Yukawa type asymptotic tails.

\section{Initial conditions.}
It is already well known that charged Hopfions can be knotted objects, \cite{fadd,battye-sutcliffe,ward,glad,ward-latt,heit,sutcliffe-good}. One of the most effective ways to create non-trivial knotted initial conditions, to be minimised, is to use the rational map ansatz technique described in \cite{sutcliffe-good}. Here the author used a degree one spherically equivariant map to compactifiy $\mathbb{R}^3 \to S_{Z_1,Z_0}^3 \subset \mathbb{C}^2$ by
\begin{equation}
 (Z_1,Z_0) = \left(\frac{(x_1+ix_2)}{r} \sin f, \cos f + \frac{i x_3}{r} \sin f \right), \label{z}
\end{equation}
where $f(0) = \pi$, $f(\infty) = 0$ and
\begin{equation}
 S^3_{Z_1,Z_0} \cong \{(Z_1,Z_0)\in \mathbb{C}^2 \mid |Z_1|^2 + |Z_0|^2 =1 \}.
\end{equation}
Using these complex coordinates an $(a,b)$-torus knot can be described as the intersection of a complex curve $q(Z_1,Z_0)$ with a unit three-sphere \cite{curve}. Hence we can formulate the rational map ansatz 
\begin{equation}
 W =\frac{\phi_1 + i\phi_2}{1+\phi_3} =\frac{l(Z_1,Z_0)}{q(Z_1,Z_0)}. \label{anz1}
\end{equation}
The inverse stereographic projection of the curve $q=0$ produces a $\phi_3=-1$ closed curve in $\mathbb{R}^3$. 
The asymptotic value of $l(Z_1,Z_0)$ in the rational map ansatz \eqref{anz1} is used to fix the boundary conditions of $\bphi$. Therefore, we need $l(r)|_{r \to \infty} = 0$ so the inverse stereographic projection gives $\bphi = (0,0,1)$ at the spacial boundary of $\mathbb{R}^3$. \\

We can now formulate an axially symmetric Hopfion initial condition as
\begin{equation}
 W= \frac{Z_1^n}{Z_1^p}, \label{Axial-ans}
\end{equation}
which has charge $Q=np$. These types of unlinked Hopfion are labelled $A_{n,p}$.
Also a Hopfion, with a position curve of an $(a,b)$-torus knot, can be formed by the rational map ansatz
\begin{equation}
W=\frac{{Z_1}^\alpha {Z_0}^{\beta}}{{Z_1}^a + {Z_0}^b} \label{ansatz}
\end{equation}
where $\alpha \in \{ x>0|x \in \mathbb{Z} \}$ and $\beta \in \{x\geqslant0|x \in \mathbb{Z}\}$. This gives a closed curve that wraps $a$ and $b$ times about the two circumferences of a torus, \cite{curve}, and has topological charge $Q=\alpha b + \beta a$ \cite{sutcliffe-good}. These tours knot configurations are labelled $K_{a,b}$. 
A torus knot with a reducible denominator can produce linked Hopfions. For example a rational map
\begin{equation}
 W= \frac{Z_1^{n+1}}{Z_1^2-Z_0^2} = \frac{Z_1^n}{2(Z_1-Z_0)}+\frac{Z_1^n}{2(Z_1+Z_0)}, \label{link-ansatz}
\end{equation}
forms two charged $n$ Hopfions linked once. We label this type of configuration as $L^{1,1}_{n,n}$. \\

This approach not only produces a non-trivial knotted location curve with an analytically known topological charge, it also gives a smooth field with the correct boundary conditions.\\
To find static Hopf solitons we set $\dot{\bphi}=0$ in the Skyrme term of \eqref{L}. This gives a non-relativistic theory which has the same static equations of motions as the those derived from \eqref{L}. This greatly simplifies the corresponding equations of motion by removing a numerically cumbersome matrix inversion. It also still facilitates time evolution by the second order dynamics derived from the sigma term. The non-relativistic equation of motion can then be numerically evolved on a discrete lattice using a fourth-order derivative approximation. We also need an additional Lagrangian multiplier, $\lambda$, to constrain $\bphi$ to take its value on $S^2$. If we periodically remove kinetic energy the potential energy will also become minimised; this will yield minimum energy static solutions. This minimisation technique produces static solutions and uses much less CPU time when compared to other similar minimisation algorithms. A numerical grid of $250 \times 250 \times 250$ points, with $\Delta x=0.08$, was found to be large enough to contain the exponentially located Hopfion. On this lattice the Hopfion can smoothly attain the vacuum value at the boundary without a noticeable expense of energy. This lattice is also fine enough not to lose the fundamental topology. The definition of the position of a Hopf soliton is sensible, but not useful for display purposes. Therefore, all the images of Hopf solitons in this paper are  plots of an isosurface of the preimage of the curve $\phi_3 = -0.85$ in the domain. This gives a surface that is a fine tube in the physical space and produces much clearer images. To show the linking number we also need to plot the preimage of a second loop, but there is no unique loop to choose. In all the plots shown we generate general loops on the target space by choosing a point, $\mbox{\textbf{g}} = (\sqrt{1-\mu^2},0,\mu)$, on $S^2$. We then find the distance on the surface of $S^2$ between $\mbox{\textbf{g}}$ and $\bphi$. This distance is then normalised by the distance between $\mbox{\textbf{g}}$ and the south pole of $S^2$, $\mbox{\textbf{e}}_{-3}$, as
\begin{equation}
\mbox{dist}(\mbox{\textbf{g}}, \bphi)= \frac{\cos^{-1}(\mbox{\textbf{g}} \cdot \bphi)}{\cos^{-1}(\mbox{\textbf{g}} \cdot \mbox{\textbf{e}}_{-3})}. \nonumber
\end{equation}
This gives many loops of constant radius on $S^2$. The preimages, of these loops, are tubes of varying thickness in the domain. Also, an isosurface of unitary value is known to intersect with the position curve of a Hopf soliton. Throughout this paper we choose $\mu = -0.9$ which is an arbitrary choice for aesthetics.

\section{$Q \leqslant 4$ trivial knots.}
\begin{figure}[!ht]
\centering
\includegraphics[width = 16cm]{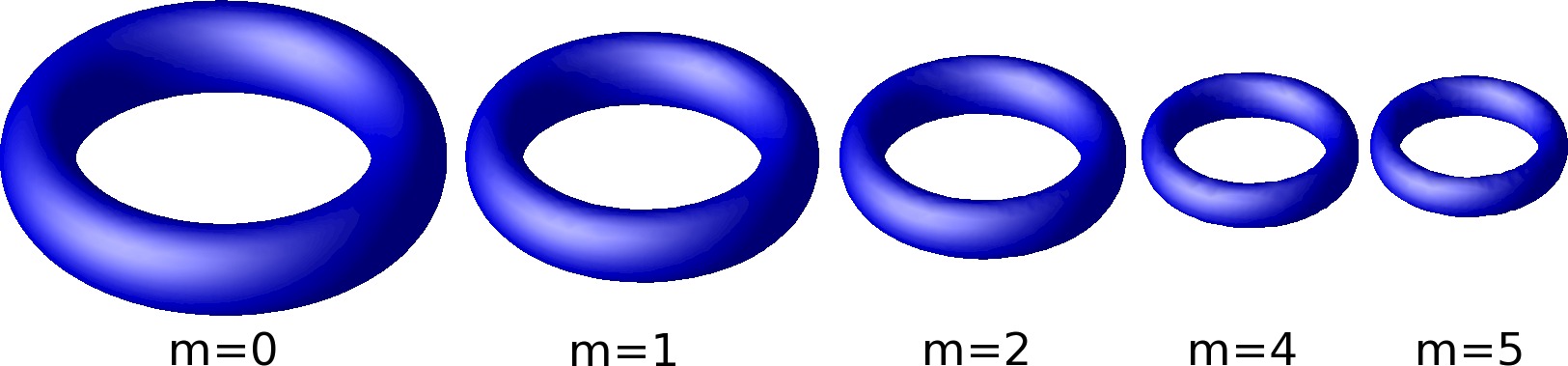}
\caption{Minimum energy topological charge one Hopfions, all on the same scale.}
\label{Q1-m=0to5}
\end{figure}

A Hopfion initial condition, where the position curve is contained completely on a plane, can be formed by the rational map ansatz \eqref{Axial-ans}. So if we set $n=p=1$ in \eqref{Axial-ans} this gives the initial condition of a topological charge one Hopf soliton which is located along a planar loop. Using this initial condition, and the above minimisation procedure, we can find the minimum energy configurations for $m=(0,1,2,4,5)$, as shown in \Fig \ref{Q1-m=0to5}.

\begin{figure}[!ht]
\centering
\includegraphics[width = 16cm]{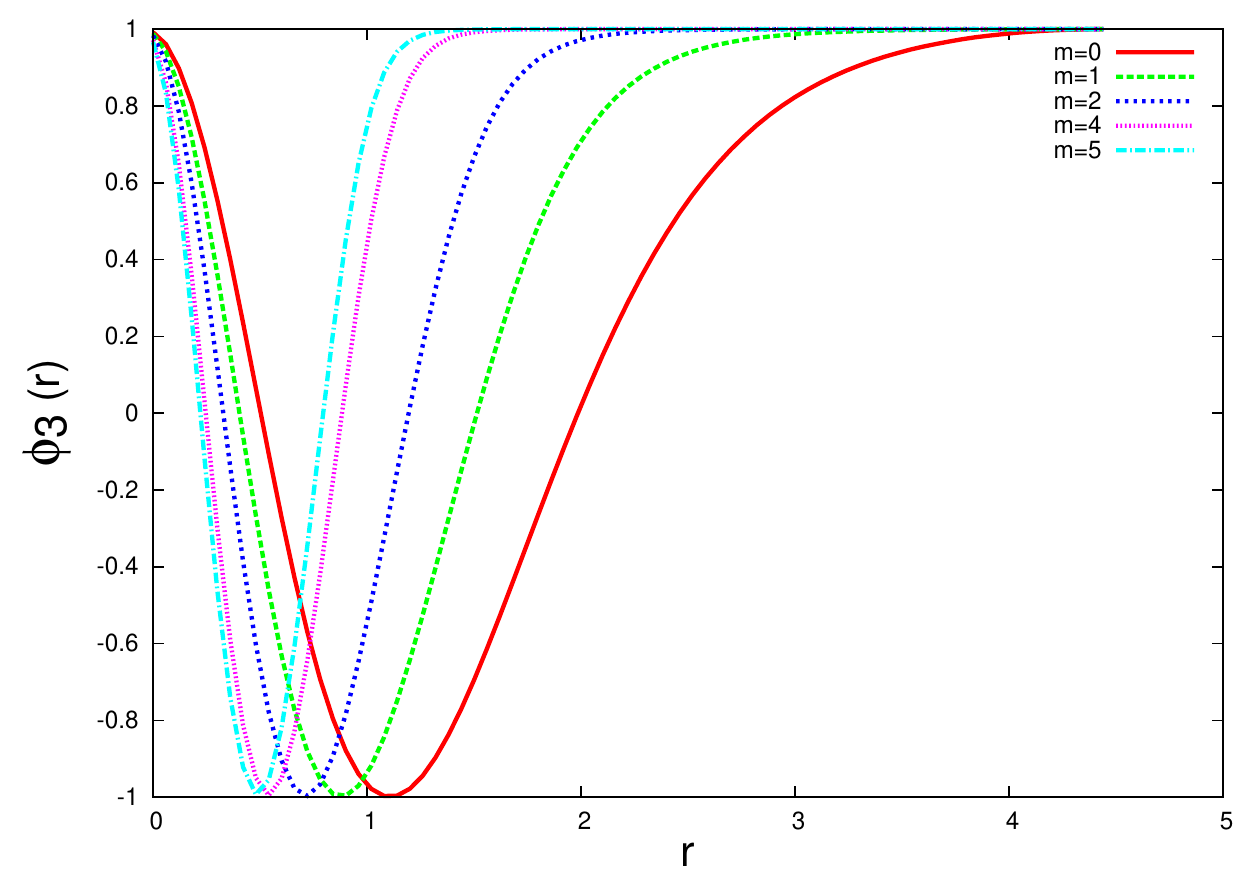}
\caption{$\phi_3$ for masses $m=0,1,2,4,5$.}
\label{slice}
\end{figure}

\begin{table}[!ht]
\centering
%\vskip -2cm
\begin{tabular}{|c|c|}
\hline
Mass, $m$ & $E$ \\
\hline 
0 & 1.236  \\
1 & 1.421  \\
2 & 1.668  \\
4 & 2.017  \\
5 & 2.170  \\
\hline
\end{tabular}
\caption{Minimum energies for $m=(0,1,2,4,5)$ topological charge one Hopfions.
}
 \label{tab-energy-h}
\end{table}

\newpage
\Fig \ref{Q1-m=0to5} shows as expected that the larger $m$ is the smaller the Hopfion is. This is due to the potential term creating an energy penalty where $\bphi =(0,0,-1)$, which corresponds to the maximum value for the potential. Thus the minimum energy Hopfion location loop becomes smaller with growing $m$. The energy for $m=0$ is $1.236$, this is within $2\%$ of the previously accepted result \cite{sutcliffe-good}. This similarity is a nice validation of our numerical procedure and the small difference can be attributed to the different choice of lattice spacing. As shown in \fg \ref{slice} for larger values of mass the field $\phi_3$ is increasingly symmetric about the Hopfion location. Due to this we have decided to perform the remaining analysis with the relatively large $m=5$. This choice of mass is arbitrary, we could have chosen $m>5$, but this choice gives sharply located Hopfions which are still large enough so the topology is not lost by the numerical lattice. We find the minimum energy $Q=1$ Hopfion to be $E_1=2.17$. \\
Using \eqref{Axial-ans} with $(n,p) =(2,1)$ this again gives a planar loop Hopfion location, but with topological charge $Q=2$. Minimising this configuration we find the minimum energy $E_2 = 3.45$, with a planar Hopfion location curve. We can construct a $Q=3$ Hopfion by setting $(n=3,p=1)$ in \eqref{Axial-ans} or $(\alpha=\beta=b=1,a=2)$ in \eqref{ansatz}. The latter configuration produces a Hopfion which is located along an unknotted twisted loop. Minimisation of these two configurations produces similar configurations with similar energy, $E_3= 4.74$, located on a twisted unknotted loop.\\

A topological charge $Q=4$ Hopfion initial condition can be made using \eqref{ansatz} with either $(\alpha =a=2,\beta=b=1)$ or $(\alpha=a=4,b=1, \beta=0)$. This gives Hopfion location curves $K_{2,1}$ and $K_{4,1}$ respectively. Minimising both of these configurations give a $\tilde{A}_{4,1}$ (twisted $A_{4,1}$) Hopfion location curve, with energy $E_4=6.051$. We can also generate $Q=4$ axial initial conditions using \eqref{Axial-ans} with $(n=p=2)$ and $(n=4,p=1)$. This gives $A_{2,2}$ and $A_{4,1}$ planar curves respectively. Minimisation of the $A_{2,2}$ configuration remains as an $A_{2,2}$, with an energy $\sim 0.6 \%$ larger than the $\tilde{A}_{4,1}$. This is within numerical accuracy of our minimisation scheme. Therefore we are not able to define which of these two configuration is the lower energy. The $A_{4,1}$ configuration also minimises to a planar curve described by $A_{4,1}$. This seems to show that twisting the loop slightly reduces the energy. The $Q=4$ planar loops are most likely long lived saddle point solutions, preserved by symmetry.
\begin{figure}[!ht]
\centering
\includegraphics[width = 16.52cm]{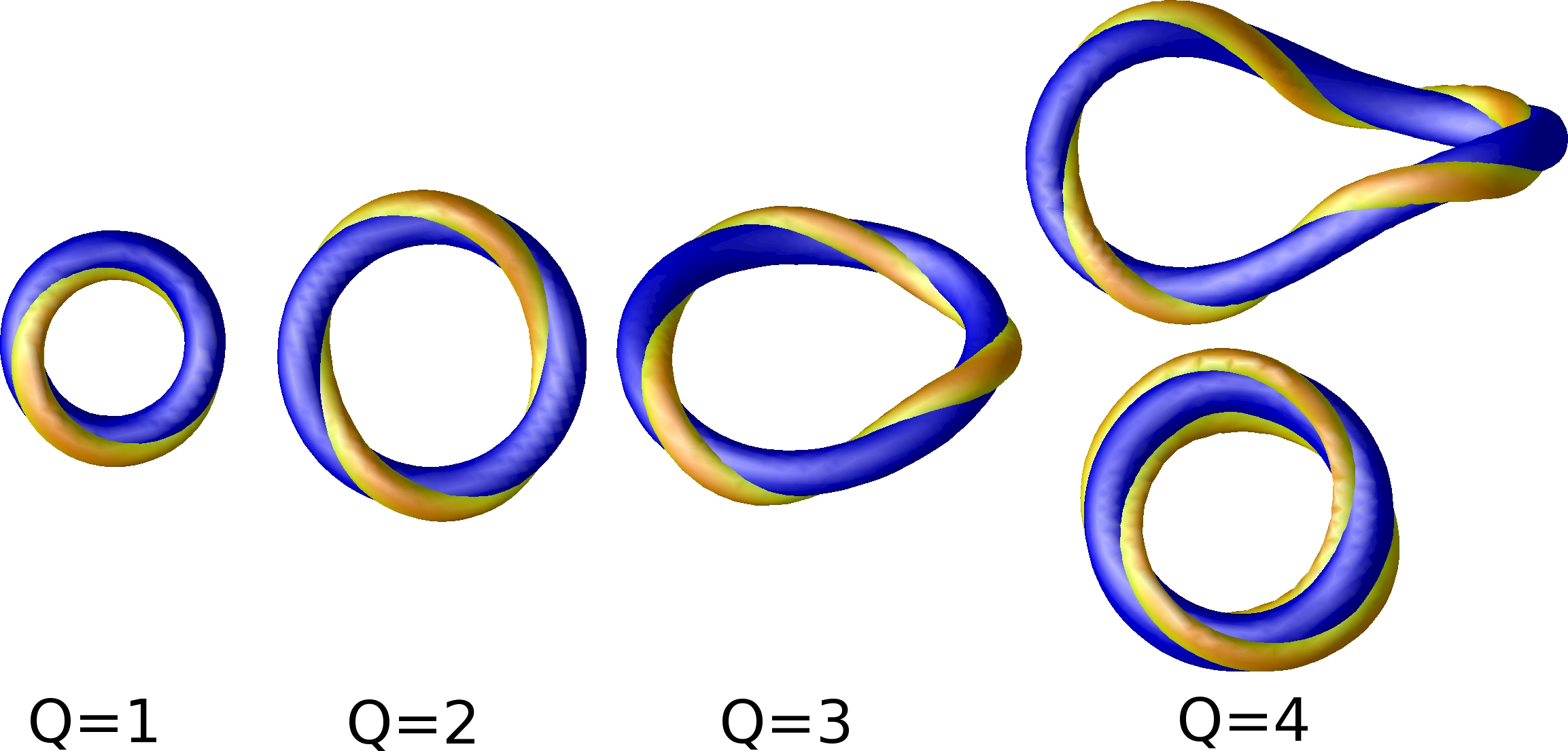}
\caption{Minimum energies for $m=5$, Hopfions.}
\label{hopf-Q=1-5-m=5-plots}
\end{figure}

\section{$Q\geqslant 5$ knotted/linked Hopfions.}
\begin{figure}[!ht]
\centering
\includegraphics[width = 15cm]{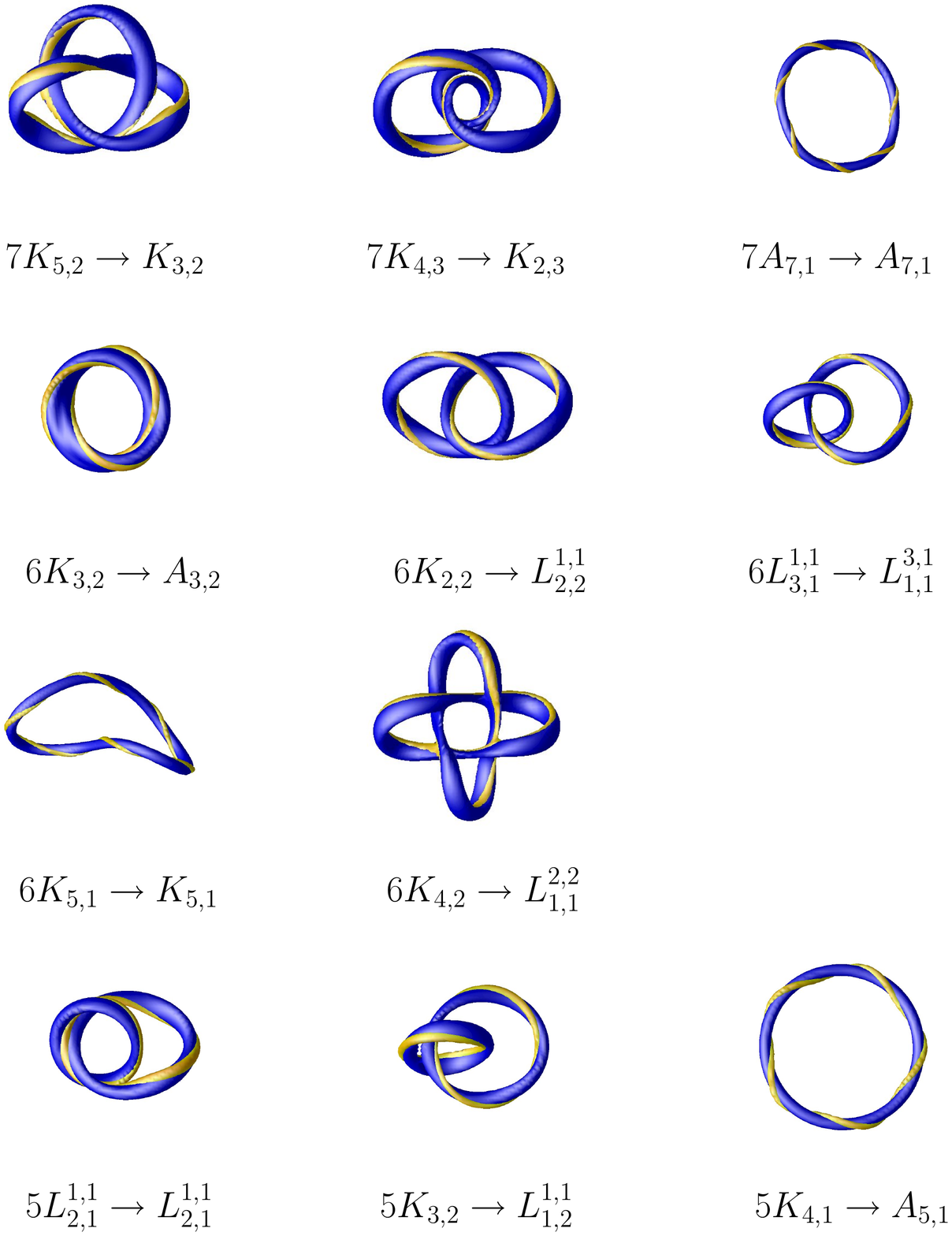}
\caption{Position curves for $m=5$ Hopfions of charge $5 \leq Q \leq 7 $. Showing their charge and linking form.}
\label{hopf-Q=5-7-m=5-plots}
\end{figure}

\begin{figure}[!ht]
\centering
\includegraphics[width = 15cm]{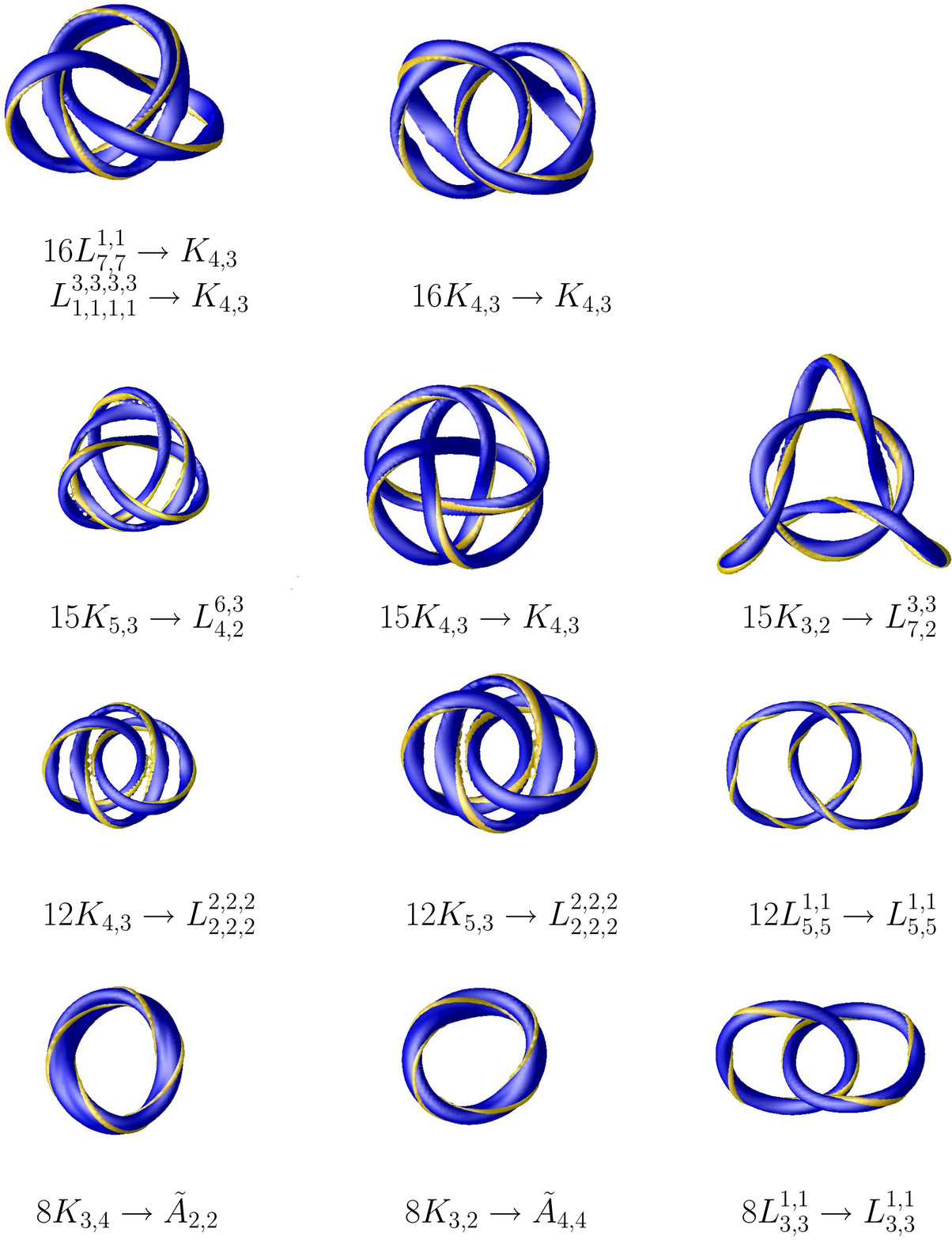}
\caption{Position curves for $m=5$ Hopfions of charge $Q=(8,12,15,16)$. Showing their charge and linking form.}
\label{hopf-Q=5-16-m=5-plots}
\end{figure}

For a select few simulations \fg \ref{hopf-Q=5-7-m=5-plots} and \ref{hopf-Q=5-16-m=5-plots} show a plot of minimum energy Hopfion locations. Also \fg \ref{hopf-Q=5-7-m=5-plots} and \ref{hopf-Q=5-16-m=5-plots} show the linking of the initial rational map ansatz and the linking of the corresponding minimum energy Hopfion curve. For topological charges $Q\leqslant 4$, both in the massive case and in the normal massless case, the Hopfion location curves are all found to be unknot solutions. For topological charges $Q\geqslant 5$ we find the minimum energy Hopfions have either a linked or knotted location curves, as shown in \fg \ref{hopf-Q=5-16-m=5-plots}. The minimum energy solutions presented have very similar qualitative features with the massless model \cite{sutcliffe-good}. The minimum energy Hopfion location curves for each charge sector have comparable linking form hence, due to the computational intensity, we have restricted our analysis to the presented charges.

\newpage

\newpage
\begin{table}[ht!]
\centering
\vskip -2cm
\begin{tabular}{|c|c|c|}
\hline
~~ & ~~$~~E~~$~~ &~~$~~E/Q^{3/4}~~$~~\\
 \hline 
$Q=4$ & ~~ &\\
$K_{2,1} \to \tilde{A}_{4,1}$&$6.05$& $2.14$\\
$K_{4,1} \to \tilde{A}_{4,1}$&$6.05$& $2.14$ \\
$A_{4,1} \to A_{4,1}$&$6.07$&$2.15$\\
$A_{2,2} \to {A}_{2,2}$&$6.09$&$2.15$\\
$L_{1,1}^{1,1} \to L_{1,1}^{1,1}$&$7.17$&$2.53$\\
\hline 
$Q=5$ & ~~ &\\
$L_{2,1}^{1,1} \to L_{1,2}^{1,1}$&$6.23$&$1.86$ \\
$K_{3,2} \to L_{1,2}^{1,1}$&$7.17$&$2.14$\\
$K_{4,1} \to A_{5,1}$&$7.43$&$2.22$\\
\hline 
$Q=6$ & ~~ & \\
$K_{3,2} \to A_{3,2}$&$8.01$&$2.09$ \\
$K_{2,2} \to L_{2,2}^{1,1}$&$8.19$&$2.14$\\
$L_{3,1}^{1,1} \to L_{3,1}^{1,1}$&$8.41$&$2.19$\\
$K_{5,1} \to K_{5,1}$&$8.60$&$2.24$\\
$K_{4,2} \to L_{1,1}^{2,2}$&$9.07$&$2.37$\\
\hline 
$Q=7$ & ~~ & \\
$K_{5,2} \to K_{3,2}$&$9.19$&$2.14$ \\
$K_{3,2} \to K_{3,2}$&$9.20$&$2.14$\\
$K_{4,3} \to K_{2,3}$&$9.64$&$2.24$\\
$A_{7,1} \to A_{7,1}$&$10.18$&$2.37$\\
\hline 
$Q=8$ & ~~ & \\
$L^{2,2}_{2,2} \to \tilde{A}_{4,2}$&$9.86$&$2.07$\\
$K_{3,2} \to A_{4,2}$&$9.88$&$2.08$\\
$K_{3,4} \to \tilde{A}_{4,2}$&$9.88$&$2.08$\\
$K_{5,2} \to A_{4,2}$&$9.97$&$2.10$\\
$L^{1,1}_{3,3} \to L^{1,1}_{3,3}$&$10.45$&$2.20$\\
\hline 
$Q=12$ & ~~ &\\
$K_{4,3} \to L^{2,2,2}_{2,2,2}$&$13.77$ &$2.16$\\
$K_{5,3} \to L^{2,2,2}_{2,2,2}$&$13.77$&$2.16$\\
$L_{5,5}^{1,1}\to L_{5,5}^{1,1}$&$15.30$&$2.37$\\
\hline 
$Q=15$ & ~~ &\\
$K_{5,3} \to L^{6,3}_{4,2}$&$16.17$&$2.12$\\
$K_{4,3} \to K_{4,3}$&$16.56$&$2.17$\\
$K_{3,2}\to L^{3,3}_{7,2}$&$17.11$&$2.25$\\
 \hline 
$Q=16$ & ~~ &\\
$L^{3,3,3,3}_{1,1,1,1} \to \tilde{K}_{4,3}$&$17.07$&$2.13$\\
$K_{4,3} \to K_{4,3}$&$17.12$ &$2.14$\\
$K_{3,2} \to K_{4,3}$&$17.25$& $2.17$\\
\hline

\end{tabular}
\caption{$m=5$ Hopfions initial and final configurations, with their respective energies.
}
\label{tab-energy-m=5}
\end{table}
 \newpage

One main difference between our results and those of \cite{sutcliffe-good} is that we are presenting potentially new energetic local minimum or saddle point solutions. As shown in \fg \ref{hopf-Q=5-16-m=5-plots} we have new linked topological charges $Q=8,12,15$ excited solutions. \\

Also our topological charge eight and six minimum energy solutions seem to have similar linking structure as in the $m=0$ case. But in the $m=5$ case the links seem to be almost on top of each other. The minimum energy location curve of the topological charge $15$ Hopfion is topologically similar to that found in \cite{sutcliffe-good}, but it is qualitatively different. \fg \ref{EoverEvsQ} shows that the energy, as a function of topological charge, has a similar growth to the $m=0$ case \cite{sutcliffe-good}.\\

\section{Infinite mass, $m\to \infty$.}
\begin{figure}[!ht]
\centering
\includegraphics[width = 15cm]{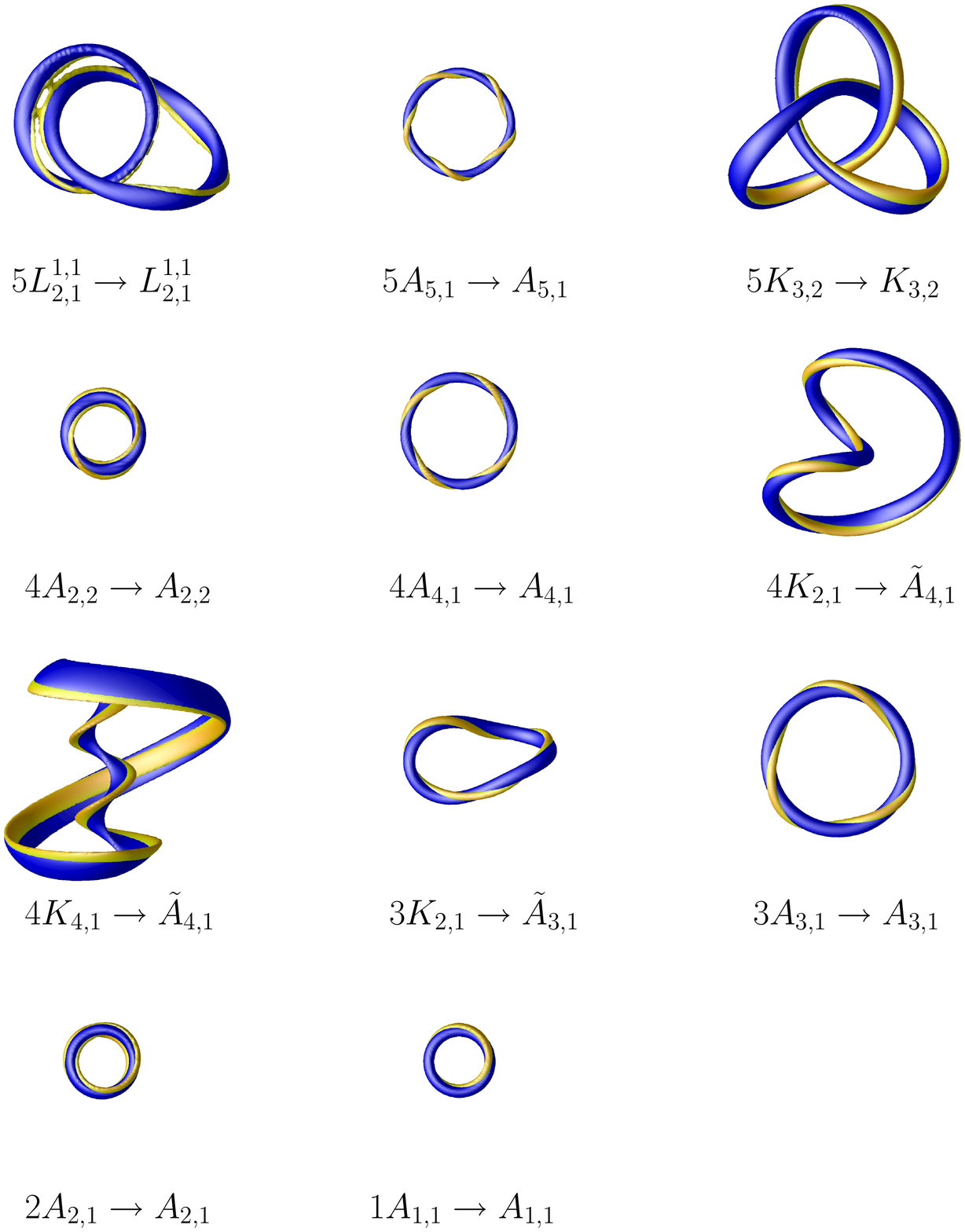}
\caption{Position curves for $m=\infty$ Hopfions of charge $1 \leqslant Q \leqslant 5$. Showing their charge and linking form.}
\label{hopf-Q=1-5-m=inf-plots}
\end{figure}

\begin{figure}[!ht]
\centering
\includegraphics[width = 15cm]{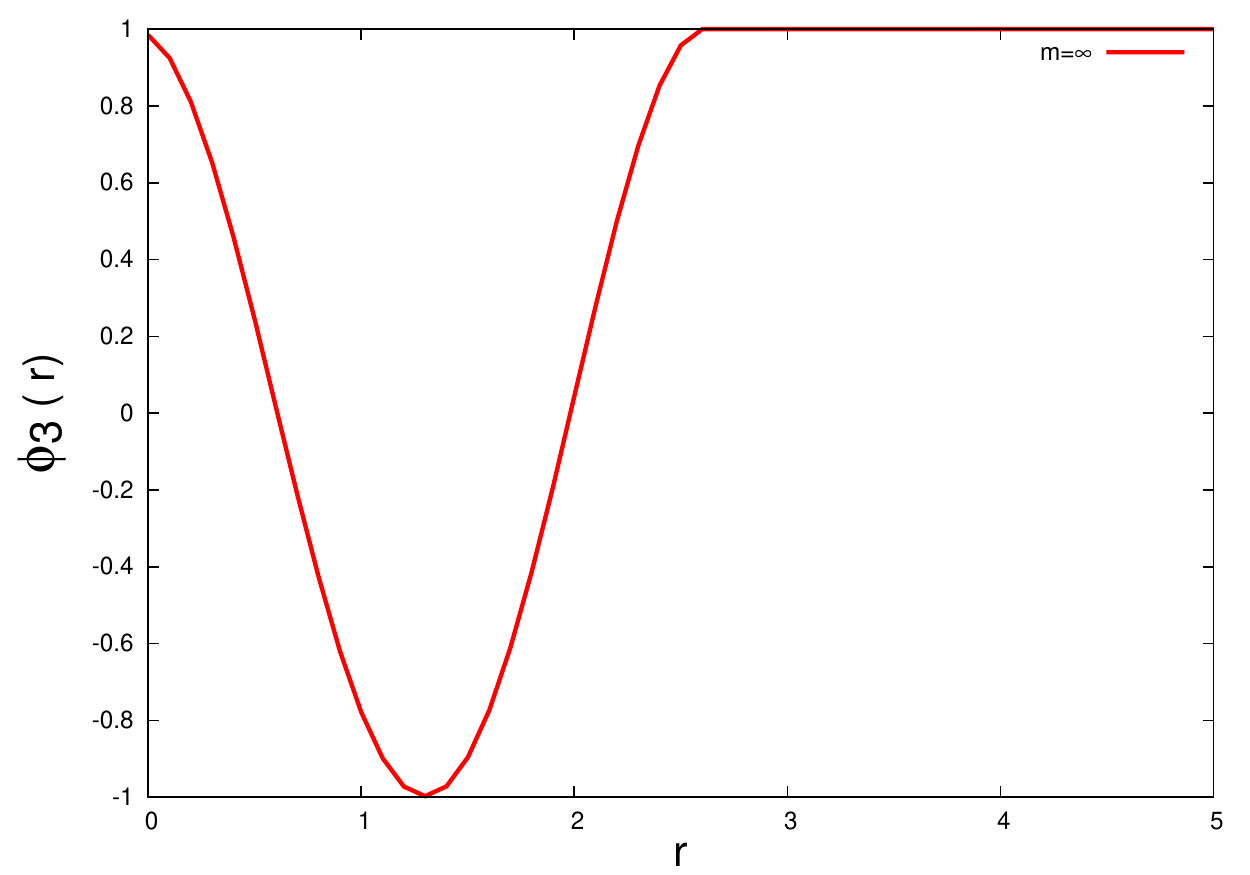}
\caption{$\phi_3$ along a radius of a $Q=1$ Hopfion, in the infinite mass case.}
\label{p3-inf}
\end{figure}

Rescaling the Lagrangian density \eqref{L} by $\bx\mapsto \sqrt{m} \bx$ gives
\begin{equation}
 \frac{\mathcal{L}(\sqrt{m} \bx)}{\sqrt{m}} =\frac{\mathcal{L}_2}{m} + \mathcal{L}_4 +\mathcal{L}_0. \nonumber
\end{equation}
Where $\mathcal{L}_a$ refers to the $a^{th}$ order derivative in the Lagrangian density. We can then define
\begin{equation}
 \mathcal{L}_{m\to \infty} = \lim_{m \to \infty} \frac{\mathcal{L}(\sqrt{m} \bx)}{\sqrt{m}} =  \mathcal{L}_4+\mathcal{L}_0. \label{inf-L}
\end{equation}
A model comprising only of a Skyrme term and a potential term has been addressed before \cite{pure-hopf}. In \cite{pure-hopf} this model was derived by setting a constant to zero. This effectively removes the sigma term in the \skf with a potential model. Soliton solutions of this model are commonly referred to as compactons \cite{pure-hopf}. This is because they reach their vacuum value in finite distance and therefore have no asymptotic tails. Hence they are effectively BPS for large separation. In order to numerically find minimum energy solutions of \eqref{inf-L} it is computationally easier to work with a modified model of the form
\begin{equation}
 \mathcal{L}_{\mbox{\begin{tiny}Modified\end{tiny}}} = \frac{\pr_0 \bphi \cdot \pr_0 \bphi}{32 \pi^2 \sqrt{2}} +\mathcal{L}_{m \to \infty}|_{\pr_t \bphi = 0}. \label{LM}
\end{equation}
This modified model can be simulated by a trivial extension of the previous \skf model. The equation of motion and the energy density of \eqref{LM} will converge with the infinite mass case, \eqref{inf-L}, in the static limit. This model, \eqref{LM}, also allows for time evolution by the second order dynamics of the purely kinetic sigma term. Again we use the rational map ansatz, \eqref{ansatz}, to give non-trivial knotted initial conditions. For this infinite mass compacton model, \eqref{inf-L}, we find the topological charge-specific minimum energy candidates in Table \ref{tab-energy-inf-m}. The numerical scheme is fundamentally the same as the one used in the finite mass case. But now due to this model having a different scale we found $\Delta x =0.1$ to be a suitable lattice spacing. The results and initial conditions of this investigation are shown in Table \ref{tab-energy-inf-m}. Also, for topological charge $1 \leqslant Q \leqslant 5$ the Hopfion location curves are shown in \fg \ref{hopf-Q=1-5-m=inf-plots}. We found that the minimum energy solutions for $1 \leqslant Q \leqslant 5$ are similar to the $m=5$ and $m=0$ models. The compact nature of this model can be seen in \fg \ref{p3-inf}; which show $\phi_3$ along a radius that is in the same plane as the planar $Q=1$ Hopfion. This shows how the compact Hopfion field attains the vacuum value in finite distance. This shows that in the $m=\infty$ model two well-separated static Hopfions do not attract or repel each other. This is due to there being no overlap of the Hopfion tails. Therefore, the string self-interaction of this $m=\infty$ model is much less than in the finite mass model. Also the shape of $\phi_3$, in \fg \ref{p3-inf}, is more symmetric about its minimum than the plots of $\phi_3$ in the finite mass model, \fg \ref{slice}. This shows that in the $m=\infty$ limit the string cross-section is much more symmetric.
Again, in this infinite mass case, we have found new local minimum energy solutions for $Q=4$; both of which have location curves described as a very twisted $A_{4,1}$. A point worth noting is that for these solutions the boundary of the numerical lattice is very far from the Hopfion. 
\begin{table}[!ht]
\centering
\begin{tabular}{|c|c|c|}
\hline
~~ & ~~$~~E~~$~~&~~$E/Q^{3/4}~~$~~ \\
 \hline 
$Q=1$ & ~~ &~~  \\
$A_{1,1} \to A_{1,1}$ & $0.86$&$0.86$  \\
\hline 
$Q=2$ & ~~ &~~ \\
 $A_{2,1} \to A_{2,1}$ & $1.37$&$0.82$  \\
\hline
$Q=3$ & ~~  &~~\\
$K_{2,1} \to \tilde{A}_{3,1}$ & $1.90$&$0.83$ \\
$A_{3,1} \to A_{3,1}$ & $2.02$&$0.90$ \\
\hline 
$Q=4$ & ~~ &~~\\
$A_{2,2} \to A_{2,2}$&$2.50$&$0.88$\\
$A_{4,1} \to A_{4,1}$ &$2.56$&$0.91$\\
$K_{2,1} \to \tilde{A}_{4,1}$&$3.19$&$1.13$\\
$L_{1,1}^{1,1} \to L_{1,1}^{1,1}$&$3.38$&$1.20$ (not shown)\\
$K_{4,1} \to A_{4,1}$&$4.22$&$1.50$\\
\hline 
$Q=5$ & ~~ &~~\\
$L_{2,1}^{1,1} \to L_{2,1}^{1,1}$ &$3.00$&$0.90$\\
$A_{5,1} \to A_{5,1}$, &$3.23$&$0.97$\\
$K_{3,2} \to K_{3,2}$ &$3.88$&$1.16$\\
\hline 
$Q=7$ & ~~ &~~\\
$k_{3,2} \to K_{3,2}$&$3.97$&$0.92$ \\
 \hline 
\end{tabular}
\caption{Infinite mass Hopfion initial and final configurations, with their respective energies.
}
 \label{tab-energy-inf-m}
\end{table}
\newpage

\begin{figure}[!ht]
\centering
\includegraphics[width = 18cm]{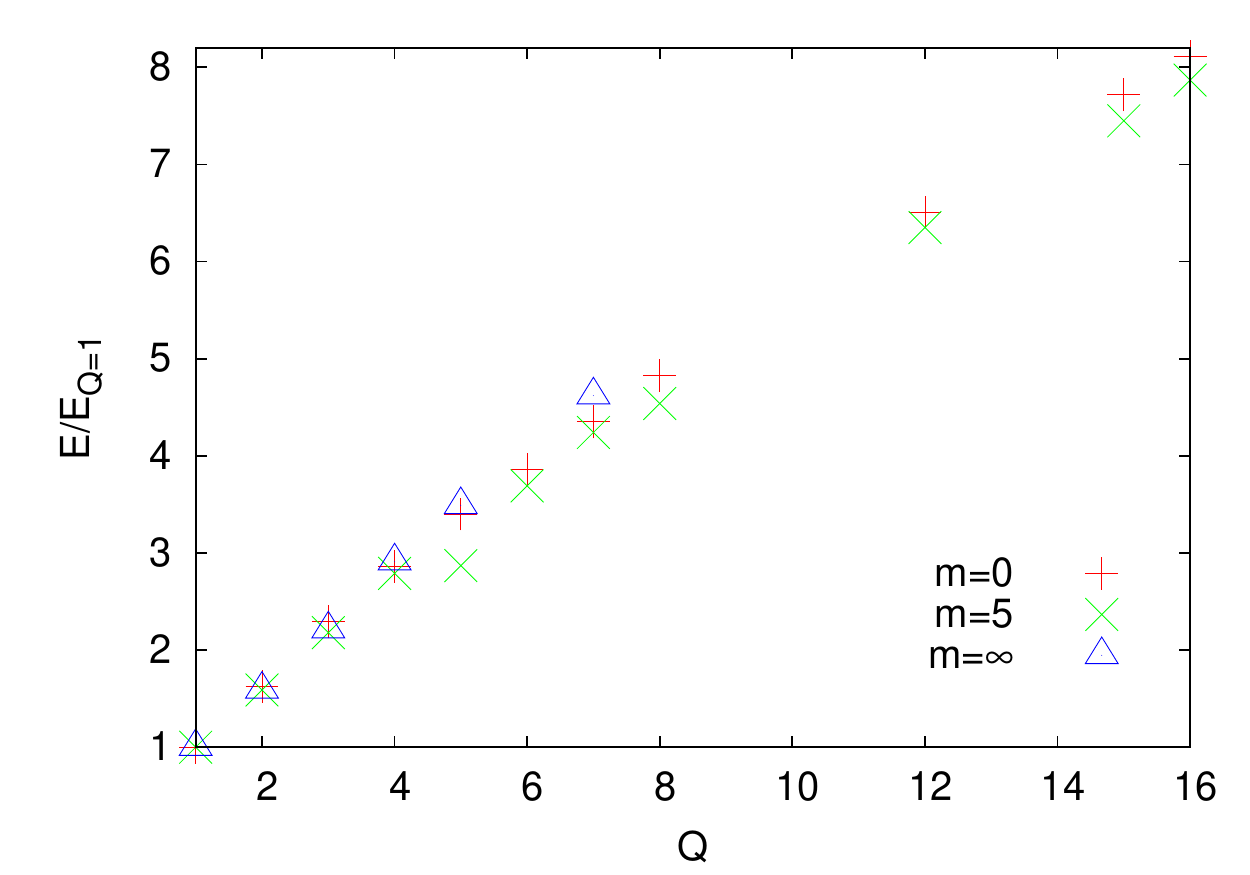}
\caption{The ratio of the energy to the topological charge one Hopfion, as a function of topological charge $Q$ for the three models.}
\label{EoverEvsQ}
\end{figure}

Table \ref{tab-energy-inf-m} shows that in this $m=\infty$ model the initial and final minimised location curves rarely differ. A good example of this is the $Q=5$ $K_{3,2}$ trefoil knot which under minimization remains as a $K_{3,2}$ trefoil knot, but in the $m=5$ and $m=0$ models minimises to a $L^{1,1}_{2,1}$. The $Q=5$, $L_{2,1}^{1,1}$, is also a lower energy solution for this $m=\infty$ model. This reluctance to deform from one location curve to another is due to the reduced self interaction of this compact Hopfion model.\\

\newpage
\begin{figure}[!ht]
\centering
\includegraphics[width = 15cm]{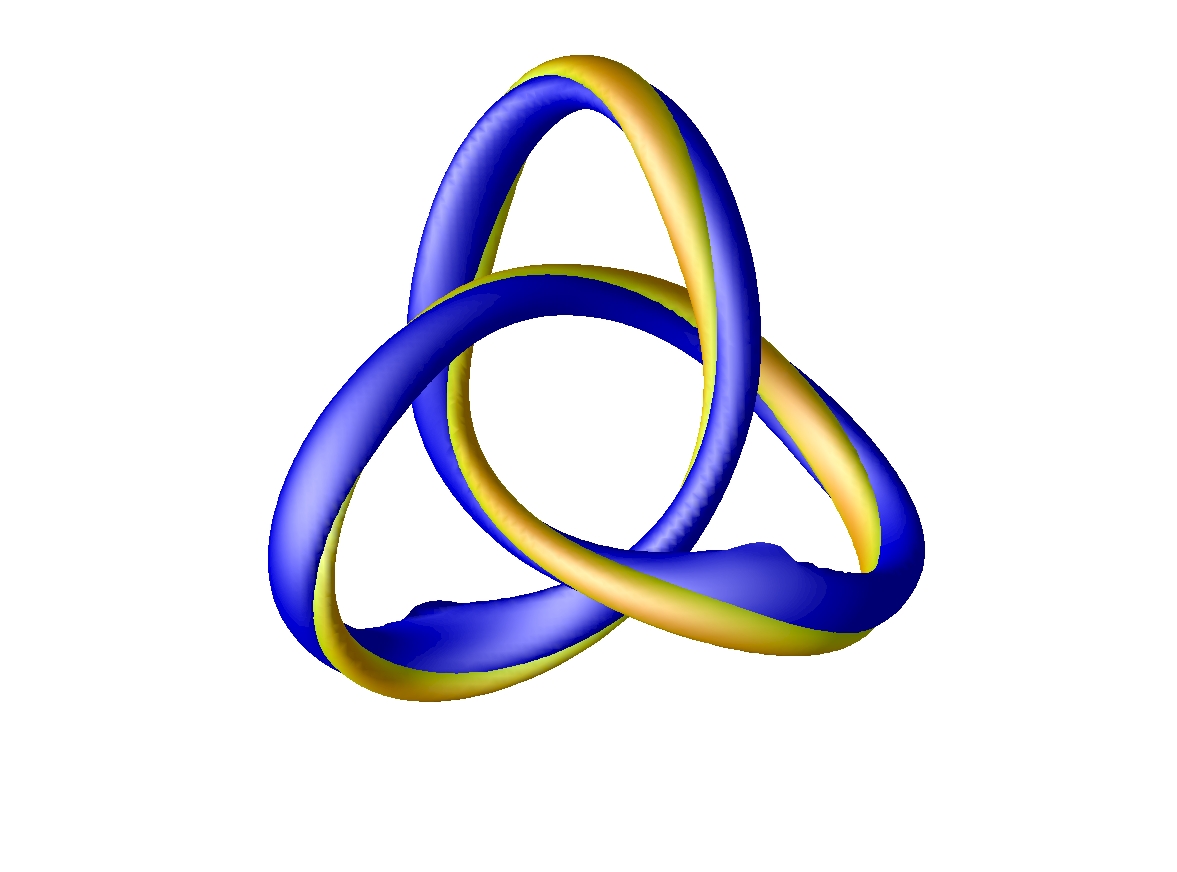}
\caption{ The $Q=7$ trefoil knot minimum energy solution, for the $m=\infty$ model.}
\label{hopf-Q=7-m=inf}
\end{figure}

We have also found a topological charge seven trefoil knot, in the $m=\infty$ model, which is shown in \fg \ref{hopf-Q=7-m=inf}. For the range of topological charges we have investigated, in this $m=\infty$ model, we have found that the charge-specific lowest energy solutions are similar to the $m=0$ and $m=5$ models. But as in the $m=5$ model we have discovered more local minima or saddle point solutions. Again, as shown in \fg \ref{EoverEvsQ}, the energy per-unit charge scales with a similar $E \sim Q^{3/4}$ grows as the $m=0$ \cite{sutcliffe-good} and $m=5$ models.

\newpage
\section{Concluding remarks.}
We have explored the Skyrme-Faddeev model with a potential term included. We found that including a potential in the model makes the Hopf solitons exponentially decay to their vacuum value. Increasing the mass makes the Hopf solitons string cross-section increasingly exponentially localized. We found for $m=5$ the minimum energy solutions are described by similar linking curves as the $m=0$ Skyrme-Faddeev model \cite{sutcliffe-good}. \\
Using a spatial rescaling we were able to formulate an infinite mass model. This infinite mass model is known to yield compact Hopfions \cite{pure-hopf}. For this infinite mass compact model we have presented a number of topologically charged solutions. We showed that the minimum energy compact solutions have similar location curves as in the usual, $m=0$, Skyrme-Faddeev model and the $m=5$ massive model. In both the $m=5$ and $m=\infty$ models we found new local minimum, or stationary point energetic solutions. The increasing localization of the strings in the two massive models reduces the string self interaction. Therefore, it is not surprising that these models possess solutions stabilised by symmetry. \\
In \fg \ref{EoverEvsQ} we showed that the $m=5$ and the $m=\infty$ models charge-specific energies seem to grow with a similar $E \sim Q^{3/4}$ trend as the $m=0$ case. Therefore, approximating Hopfions as Kirchhoff rods \cite{kirchhoff-rods} could also be successful in the massive model.\\
\\
\textbf{Acknowledgement.} I would like to thank Professor P.M. Sutcliffe and Doctor D. Harland for their useful discussion. I would also like to thank the Department of Mathematical Sciences at Durham university where a significant amount of the presented research was performed.

\end{document}